\documentclass{elsart}
 \usepackage{graphicx}
\usepackage{amsmath}
\usepackage{amssymb}          %Simbolos de AMS.
\usepackage{amsfonts}         %Fonts de AMS.
\begin{document}
\begin{frontmatter}

\title{Dynamical localization for Bloch electrons in Magnetic and Electric Fields}

\author{Manuel Torres}
 \ead{torres@fisica.unam.mx}
\address{Instituto de F\'isica, Universidad Nacional Aut\'onoma
                          de M\'exico, Apdo. Postal 20-364, M\'exico D.F. 01000,
                          M\'exico}

\author{Alejandro  Kunold}
 \ead{akb@correo.azc.uam.mx}
\address{Departamento de Ciencias B\'asicas, Universidad Aut\'onoma
             Metropolitana-Azcapotzalco, Av. S. Pablo 180,
             M\'exico D.F. 02200, M\'exico}

\begin{abstract}
We study the time evolution of wave packets of noninteracting electrons  in a  two-dimensional periodic  system in the
presence of magnetic and electric fields. The model includes  consistently the coupling between Landau levels as well as the 
periodic and external field contributions. 
It is shown that the electric  field produces localization along its 
longitudinal direction, this  effect  is a physical consequence  of the  quantization of the energy levels with respect to the longitudinal quasi-momentum.  
For incommensurate orientations of $\mathbf{E}$  relative to the  lattice the wave packet becomes  localized  
 in both direction.
  \end{abstract}

\end{frontmatter}
%\section{Introduction}
Localization of waves in quasiperiodic system has been the subject of great interest in condensed matter and related areas in physics. 
 An electron subject to a sum of potentials which are periodic but incommensurate with one another represents a case  that is intermediate between a random and a perfectly ordered system. In an incommensurate situation there can be  both  localized and extended states, with the intermediate  possibility of a metal-insulator transition as the energy and-or the strength of the potential are varied. Aubry and Andre \cite{Aubry1} presented a one-dimensional  model that  exhibits  this behavior;
 the dynamics of their model is governed by  the Harper \cite{Harper1}.
 The same model  also arises in the the study of  an electron moving in a two-dimensional lattice  subject to 
a magnetic field, both in the limiting cases of either a very large \cite{Harper1}  or a very small \cite{Rauh1} potential strength compared to the magnetic field intensity.  The problem of electrons moving under the  simultaneous influence of a periodic
potential and  a magnetic field has been discussed by many
authors \cite{Harper1,Rauh1,Peierls1,Zak2}; the 
spectrum displays an amazing complexity including various kinds of
scaling and a Cantor set structure \cite{Hofta1}. The addition of an electric field leads to  interesting new phenomena 
that makes  its analysis worthwhile.  For example, as the strength of the electric field increases  the longitudinal quasi-momentum is quantized leading  to the appearance of a "magnetic Stark ladder" in which the bands are replaced by a series of quasi discreet levels \cite{Kunold1,Torres1}.  The object of the present work is to study the time evolution   of electron wave packets  in a periodic 
two dimensional system subject to external electric  and magnetic fields.  Previous studies of  this problem were   usually  restricted to a  single band limit derived from a tight binding  and/or  the Peirles  approximations \cite{Brito1,Claro1}. 
Here we presents a model which   consistently includes  the coupling between Landau levels as well as the 
periodic and external field contributions. This treatment becomes essential if one would like to analyze    physical realistic  situations. 
The study of these phenomena has become experimentally accessible with the recent developments in the fabrication of antidot
arrays in lateral superlattices by ion beam and atomic force lithography  \cite{Esslin3,Klit4}. For example,   
the observation of the Hofstadter Butterfly spectrum through
the measurement of the magnetoresistance and Hall conductance in
artificial arrays of anti quantum dots has just been achieved \cite{Klit2}.

%\section{The Model}
We consider the motion of an 
 electron in a two-dimensional periodic potential $V\left(x,y\right)$
subject to a uniform magnetic  field  $\mathbf{B}$  perpendicular to the plane and to a constant electric field $\mathbf{E}$, lying 
on the plane  according to $\mathbf{E} = E \, (cos \theta, sin \theta)$, with $\theta$ the angle between $\mathbf{E}$
 and the lattice $x-$axis. 
The dynamic is governed by the  Schr\"odinger equation 
\begin{equation}
i\hbar \frac{\partial}{\partial t}   \psi
=\left[\frac{1}{2m}\left(\mathbf{p}+e\mathbf{A}\right)^2
-e\phi\left(x,y\right)+V\left(x,y\right)\right]  \psi,
\end{equation}
the vector and scalar potentials  are selected to yield: $\nabla \times \mathbf{A}=B \hat{k}$,  and  $\mathbf{E}=-\nabla \phi -\partial \mathbf{A}/\partial t$. 
For simplicity we shall consider a periodic square lattice: $V\left(x,y\right)=U_0\left[\cos\left(2\pi  x/a \right)
+\lambda\cos\left(2\pi y/a\right)\right].$  Two characteristic parameters of the system, the cyclotron frequency 
$\omega_c = e B /m$ and the magnetic length $l_B = \sqrt{\hbar /(e B)}$ set the reference scales for  energy and lengths respectively. 
Following the formalism developed in references \cite{Kunold1,Torres1}, we introduce a canonical transformation to new variables 
$Q,P$,  acoording to 
\begin{align}
Q_0&=t \, ,  &P_0 = i \hbar \partial_t+e \tilde{\phi}+\frac{m}{2} \left( \vert \mathbf{E} \vert /B  \right)^2  ,  \nonumber \\
eQ_1&=p_y+eA_y+  m  E_x / B   \, , &P_1 =p_x+A_x - m E_y /B  ,\nonumber \\
eQ_2&=p_x+ e\tilde{A}_x-E_x t \, , &P_2=p_y+ e\tilde{A}_y-E_yt,
\label{aab}
\end{align}
where $\tilde{\phi}=\phi+\mathbf{r}\cdot \mathbf{E}$,
$\tilde{A}_x=A_x+By+E_xt$ and $\tilde{A}_y=A_y+Bx+E_yt$.  It is easily verified that the transformation is indeed  canonical, 
the  new variables obey  the commutation rules: $\left[Q_0,P_0\right]=-i \hbar$, and $\left[Q_1,P_1\right]= \left[Q_2,P_2\right]=i \hbar B$; all other commutators being zero.
The inverse transformation gives $x= l_B^2 \left(eQ_1-P_2 \right)/\hbar $ and $y=l_B^2 \left(e Q_2-P_1 \right)/\hbar$. 
The operators $(P_0,  Q_2 , P_2)$ can  be  identified with the generators of the electric-magnetric translation simetries \cite{Ashby1,Kunold1}.
Final results are independent  of the selected gauge. 
From the operators in Eq. (\ref{aab}) we construct two pairs of
harmonic oscillator-like ladder operators: $(a_1,a_1^{\dag})$, and $(a_2,a_2^{\dag})$ with :
\begin{equation}\label{opasc}
a_1 = \sqrt{\frac{1}{2 \hbar B}  }  \left(P_1-iQ_1\right), \qquad   a_2 = \sqrt{\frac{1}{2 \hbar B}  }  \left(P_2 -i Q_2\right), 
\end{equation}
obeying:  $[a_1,a_1^{\dag}]=[a_2,a_2^{\dag}]=1$, and $[a_1,a_2]=[a_1,a_2^{\dag}]=0$.
Using these  variables, the Schr\"odinger equation
can be recast as $P_0\left\vert \psi \right\rangle=H\left\vert \psi \right\rangle$
with  the Hamiltonian  given by
\begin{equation}\label{aac}
H= \hbar \omega_c \left( \frac{1}{2} + a_1^{\dag} a_1 \right) + V
-e l_B \left( \tilde{E}a_2 + \tilde{E}^{*}a_2^{\dag} \right),
\end{equation}
here $\tilde{E}=(E_x+iE_y)/\sqrt{2}$, and  $V(x,y) \equiv V(Q_1 - P_2, Q_2 - P_1)$ has to be  expressed in terms of the ladder operators using Eq. (\ref{opasc}).  The properties of the system depends on five dimensionless parameters:
(i) $\lambda$ modulates the relative potential  amplitudes in $x$ and $y$ directions,
 $\lambda >1$ ($\lambda < 1$) corresponds to a weaker modulation along the $x$ ($y$) direction.
 (ii)  $\sigma^{-1}= \phi / \phi_0  = a^2 B/(h/e)$ is  the number of flux quanta in a unit cell of area $a^2$.
 (iii) $\theta$ is the relative  angle between the electric field direction and 
the lattice $x-$axis. The two other parameters  $K$ and $\rho$ are the potential strength in units of
 $\hbar \, \omega_c$, and the electric field in units of $U_0/(e a)$, hence: (iv) $K=U_0/\hbar \, \omega_c$, and  (v)  $\rho = e a \vert \vec E \vert  / U_0$, respectively.
As an appropriate set of basis functions we choose 
the eigenstates of the number operators $a_1^{\dag}a_1$
and $a_2^{\dag}a_2$ 
\begin{align}
a_1^{\dag}a_1\left\vert \mu,\nu\right\rangle
&=\mu\left\vert \mu,\nu\right\rangle,&
a_2^{\dag}a_2\left\vert \mu,\nu\right\rangle
&=\nu\left\vert \mu,\nu\right\rangle,
\label{mor}
\end{align}
where $\mu$ labels the Landau levels and
$\left\vert \mu,\nu \right\rangle=\left\vert \mu \right\rangle
\otimes\left\vert \nu \right\rangle$.
These states  are also
eigenvectors of the angular momentum 
operator $\mathcal{J}=(P_1^2+Q_1^2-P_2^2-Q_2^2)/2$:
$ \mathcal{J}\left\vert \mu,\nu \right\rangle
= \hbar \left(\mu-\nu\right)
\left\vert \mu,\nu \right\rangle. $
In the coordinate representation the wave function for this  state takes the form 
\begin{equation}\label{profi}
\psi_{\mu,\nu}\left(x,y\right)
=\frac{e^{-\left(x^2+y^2\right)/4l_B^2}}{\sqrt{2\pi}}
i^{\nu} \sqrt{\frac{2^{\mu} \mu!}{2^{\nu}\nu!}}
\left(\frac{x+iy}{l_B}\right)^{\nu-\mu}
L_{\mu}^{\nu-\mu}
\left[\frac{\left(x^2+y^2\right)}{2 l_B^2}\right],
\end{equation}
where $L_{\mu}^{\nu-\mu}$ are the generalized Laguerre polynomials, the previous formula is valid for $\mu \le \nu$, if 
$\mu \ge \nu$ the indices $\mu$ and $\nu$ must be exchanged on the right hand side.
These wave functions are centered in
the origin and its   mean square  radius 
is given by $\left\langle\mu,\nu\left\vert r^2\right\vert\mu,\nu \right\rangle=
8 \, l_B^2 \, \left(\mu+\nu+1\right).$
The initial form of the wave packet has been selected as the lowest
Landau level with zero angular momentum:  $\psi_{0,0}\left(x,y\right)$.
The  evolution of the system is obtained applying
the  operator
$O\left(t\right)=\exp\left(-itH\right)$ to the initial state $i.e.$ $ \Psi\left(t,x,y\right)=\left\langle x,y\left\vert
O\left(t\right)\right\vert 0,0\right\rangle.$
The Hamiltonian in  (\ref{aac}) is represented by a matrix   $\mathbb{H}$ with elements evaluated in 
the base given in Eq. (\ref{mor}), the matrix elements can be worked out as 
\begin{multline}\label{ele}
\left\langle \mu^{\prime}, \nu^{\prime} \left\vert H
\right\vert \mu,\nu \right\rangle  =\\
\delta_{\mu^{\prime},\mu}
\bigg[ \hbar \omega_c \left(\frac{1}{2} + \mu\right)
\delta_{\nu^{\prime},\nu} 
-e l_B \left( \tilde{E}\sqrt{\nu^{\prime}}\delta_{\nu^{\prime},\nu+1} 
 + \tilde{E}^*\sqrt{\nu}\delta_{\nu^{\prime}+1,\nu} \right) \bigg] + \\
U_0 \bigg[ D_{\mu^{\prime},\mu}\left(\sqrt{\pi\sigma}\right)
D_{\nu^{\prime},\nu}\left(-i\sqrt{\pi\sigma}\right) 
+  \lambda D_{\mu^{\prime},\mu}\left(-i\sqrt{\pi\sigma}\right)
D_{\nu^{\prime},\nu}\left(\sqrt{\pi\sigma}\right)+\mathrm{c.c.} \bigg],  \\
\end{multline}
%
%FIGURA
\begin{figure}
\includegraphics[width=130mm]{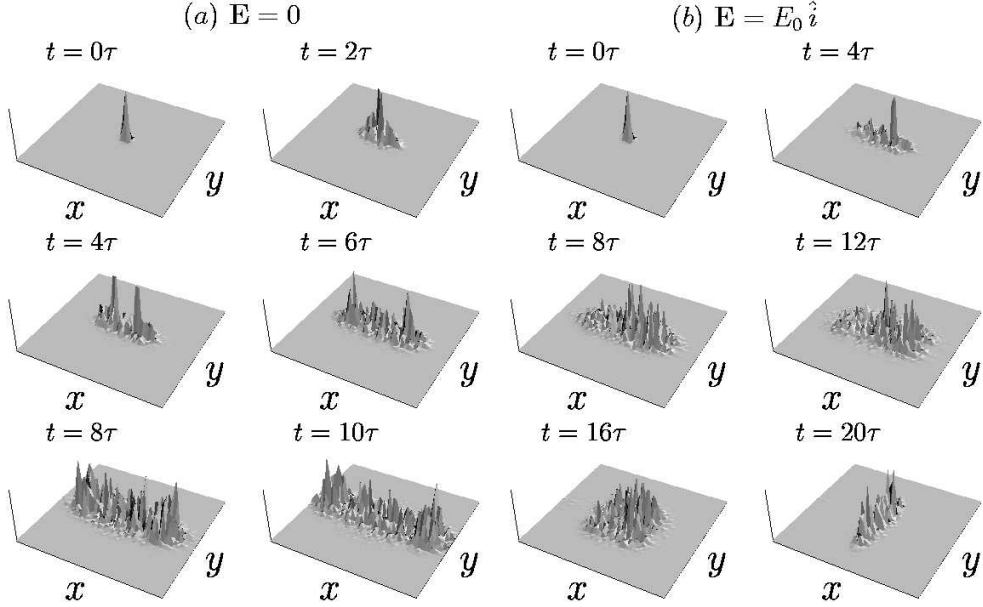}% Here is how to import EPS art
\caption{\label{figure1} 
Time evolution of an initially localized wave packet in a $2D$ periodic asymmetric 
potential ($\lambda = 2$; lower modulation amplitude in the $x$ direction)  for:  (a)  $\mathbf{E}=0$, 
and (b) an electric field   pointing  in the  $x$ direction, with an intensity   $\rho = e a \vert \vec E \vert  / U_0 =0.1$. 
Time is measured in units of $\tau = 2\pi/ \omega_c$. Here and throughout the paper the  value of the other parameters are $\sigma=1/2$, $K = 5.5$. }
\end{figure}
\noindent
where $D_{\mu^{\prime},\mu}\left(z\right) =
\left\langle\mu^{\prime}\left\vert e^{za_1-z^*a^{\dag}_1}
\right\vert\mu\right\rangle   =\psi_{\mu^{\prime},\mu}
\left(\sqrt{2}\mathrm{Re}z,\sqrt{2}\mathrm{Im}z\right)$
are the matrix elements of the coherent state operator
$e^{(za_1-z^*a^{\dag}_1)}$, they can be expressed using the same functions 
that determine  the wave function profiles (Eq. \ref{profi}).
The Hamiltonian in (\ref{ele}) is  diagonalized numerically  by the unitary matrices  $\mathbb{U}$ 
that are constructed from   the column eigenvectors of $\mathbb{H}$.
In this base the evolution operator takes the form 
$\mathbb{O}\left(t\right)=\mathbb{U}^{\dag}
\exp{\left(-it\mathbb{E}\right)}\mathbb{U}, $
where $\mathbb{E}$ is the diagonal matrix containing the energy eigenvalues.
The time evolution of the system is given by
\begin{equation}\label{eva}
\Psi\left(t,x,y\right)=
\sum_{\mu,\nu=0}^{\infty}\mathbb{C}^{\mu}_{\nu}\left(t\right)
\psi_{\mu,\nu}\left(x,y\right),
\end{equation}
where $\mathbb{C}\left(t\right)=
\mathbb{O}\left(t\right)\mathbb{C}\left(0\right)$, with the initial state
$\mathbb{C}^{\mu}_{\nu}\left(0\right)=\delta_{\mu,0}\delta_{\nu,0}$.
A global characterization of the dynamic evolution of a wave packet is provided by 
displaying  $3 \,D$ graphics of  the  probability density $ \rho\left(t,x,y\right)
=\left\vert\Psi\left(t,x,y\right) \right\vert^2,$
as a function of the space-time variables. The  localization properties of the system 
are characterized by the evolution  of the mean square displacement
 $\sigma^2_r (t)$. Both the electric field and the  $\lambda $ parameter can induce anisotropy 
transport properties, so its is sometimes convenient to separately refer to the  $x$ and $y$ variance:
$ \sigma_x^2\left(t\right) 
=\left\langle x^2 \right\rangle_t
- \left\langle x \right\rangle^2_t, $ and
$\sigma_y^2\left(t\right)
=\left\langle y^2 \right\rangle_t
- \left\langle y \right\rangle^2_t,$
 with $\sigma^2_r = \sigma^2_ x + \sigma^2_y$. 
As an additional criteria we consider the   
 Shannon information  entropy\cite{Shannon1}, defined by 
\begin{equation}
S\left(t\right)=- \int  \rho(t,\vec r ) \,\,  \ln \rho(t,\vec r ) \,\, d^2r.
\end{equation}
This quantity  characterize  the spreading of the probability density distribution, it provides a 
 measure of the spatial delocalization of the wave packet.  The lower this quantity is, the more
localized the electron is. 
%
%FIGURA
\begin{figure}
\includegraphics[width=130mm]{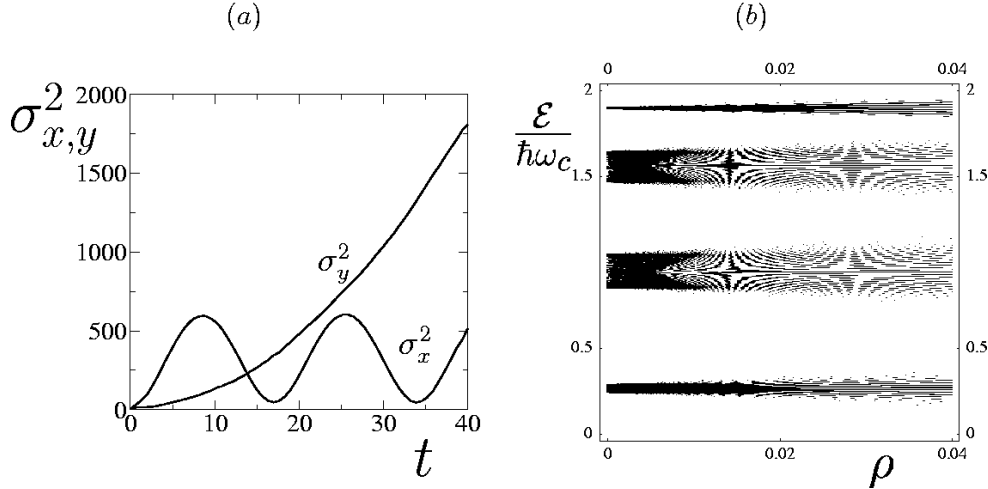}% Here is how to import EPS art
\caption{\label{figure2} 
(a) Time evolution of the $\sigma_x^2$ and $\sigma_y^2$ variances for the wave packet 
represented in Fig.~\ref{figure1}b,  the wave packet  is dynamically  localized in the longitudinal direction and    spreads  ballistically in the direction of stronger modulation $\sigma_y^2 \sim t^2$.
(b) Energy subspectra as a function of the parameter $\rho$ 
($\rho = e a \vert \vec E \vert  / U_0$). The  transverse quasimomentum is fixed to $k_y =0$ and  the longitudinal quasimomentum takes all possible values in the  first Brillouin zone, $k_x \in (-\pi/2 a, + \pi /2 a)$. The first two Landau levels are shown,  $\sigma =1 /2$, so  every Landau level splits in two bands. A transition from  extended  to localized states is observed as the electric field intensity is increased. }
\end{figure}

%
%\section{Results and Discussion}
In what follows we analyze  the propagation of wave packets and its properties for some specific examples. Using the  Peirls tight-binding model Nazareno and Brito \cite{Brito1} studied the electron wave propagation, they  found that for $\sigma$ rational the electron propagates ballistically, whereas for  irrational $\sigma$ the wave remains localized. Using our   model that  incorporates  inter-Landau couplings,  we have verified that these results remain valid; details are not presented  due to space   limitations.  In what follows we shall restrict to the case of $\sigma = 1/2$ and fix  the  parameter  $K$   to $K=5.5$.  The Hamiltonian in (\ref{ele}) is diagonalized   numerically 
considering  a base  of $\mu = 4$ landau levels and $\nu =850$ excitations, this selection proves to give 
good convergence. This was verified by checking  that the participation of states with higher values 
of $\mu$ and $\nu$ remains smaller than  one part in $10^7$. 
%

%
%\subsection{Directional localization induced by the lattice  anisotropy and the electric field.}
{\bf Directional localization induced by the lattice  anisotropy and the electric field.} 
If the lattice potential is anisotropic ($\lambda \ne 1$)  electron  transport is 
expected along the direction of weaker modulation.  
Fig.~\ref{figure1}a presents the time evolution of a wave packet initially localized, in the absence 
of electric fields, and for a  lower modulation along the $x$-direction ($\lambda =2$). Clearly as expected,  
ballistic transport takes place along the direction of lower modulation.  If an electric field  in the $y$ direction is  applied, the characteristic properties of ballistic transport along the $x$-axis would remain valid. Instead, let us consider that an electric 
 field pointing in the direction of lower modulation ($x$-axis) is switched-on, for sufficiently strong field a transition takes place and the electrons are re-directed into the direction of stronger modulation and simultaneously  localized in the direction of lower modulation (x-axis). This behavior is displayed in   Fig.~\ref{figure1}b,  we observe that initially the effect of the lower potential modulation   dominates and the   wave package spreads in this  direction  until it reaches a maximum
size (at a time $ t \approx 10 \, \tau$,  $\tau = 2\pi / \omega_c$), after this time the diffusions  along the longitudinal directions stops, with the   mean square  displacement $\sigma^2_x$    showing  strong oscillations; simultaneously   the package  spreads
along the transverse direction ($y$ axis). This behavior is corroborated by   the behavior of the $x$ and $y$ variance.  In Fig.~\ref{figure2}a.  $\sigma_x^2$ and $\sigma_y^2$ are plotted as function of time, the evolution in the transverse higher modulation direction is ballistic, whereas  localization in the longitudinal direction  with a strong oscillatory behavior for $\sigma^2_x$ is observed.  Hence the system has undergone  a metal-insulator transition in the $x$ (longitudinal) direction and a insulator-metal transition in the $y$ (transverse) direction.  The high  amplitude of the   $\sigma^2_x$ oscillations  have its origin in the competing contributions between the  modulation of the periodic potential and the electric field. The change in the direction of 
 transport, forcing the electron to overcome  the stronger  potential modulation,  is related to the behavior  of the spectrum and eigenfunctions of the  Hamiltonian (\ref{ele}). Fig.~\ref{figure2}b presents the energy  subspectra as a function of the electric field strength for the first two Landau levels, notice   that for  $\sigma =1 /2$ every Landau level splits in two bands.  To generate the subespectra, the transverse quasimomentum is fixed to a value $k_y =0$, while the longitudinal quasimomentum takes all possible values in the  first Brillouin zone, $k_x \in (-\pi/2 a, + \pi /2 a)$.  It is observed a transition from wide to extremely narrow minibands, simultaneously the  eigenfunctions change from extended to localized  with respect to the longitudinal quasi-momentum. This extremely narrow  band  corresponds to the generation of quasi discreet levels or "magnetic Stark ladder" \cite{Kunold1,Torres1}, and leads to the longitudinal localization. 
 
 A phenomena similar to the one presented here,  has been reported by  
 Ketzmerick, Kruse and Springsguth \cite{Ketz5}. In that  case, the   transition that induces  transport along the direction of stronger modulation and localization in the direction of  weaker modulation is related to avoided band crossing  that produces transitions from band to discreet levels or vice versa as the  strength of the periodic potential is increased. In our present example the transition is related to the change from  wide to extremely narrow minibands 
induced by  the electric field.

%\subsection{Commensurate vs. incommensurate directions of the electric field}
{\bf Commensurate vs. incommensurate directions of the electric field}
Finally, we consider the effects produced by  the commensurability of
the electric field direction.  We refer  to a commensurate direction
if the electric field is oriented in such that  a way that the angle $\theta$ fulfills the condition $tan \, \theta = m_2/m_1$, with 
$m_1$ and $m_2$ relatively prime integers, this condition ensures that a spatial periodicity is preserved 
along the longitudinal and transverse direction of $\mathbf{E}$.  In the rest of the paper we shall restrict to a  symmetric lattice $(\lambda = 1)$.
%
%FIGURA
\begin{figure}
\includegraphics[width=130mm]{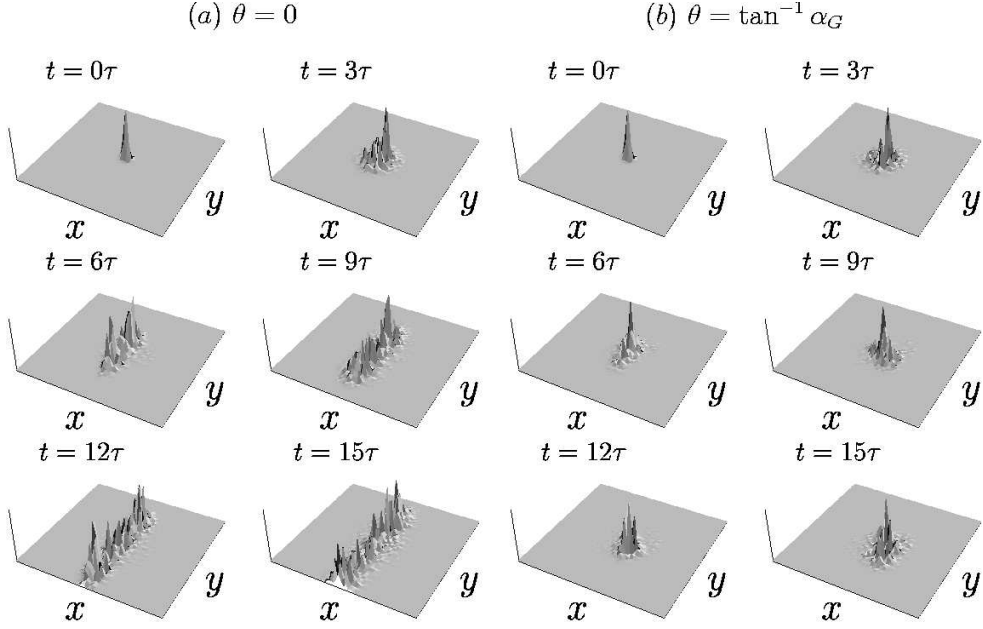}% Here is how to import EPS art
\caption{\label{figure3} 
(a) Wave packet evolution when the electric field is oriented in an
 commensurate direction $\theta=0$, the electron is  localized in the longitudinal direction.
(b) Wave packet evolution when the electric field is oriented in an
 incommensurate direction $\theta=\tan^{-1}\alpha_G,$  $\alpha_G=\left[(\sqrt{5}-1)/2\right]$.
The wave packet is localized in both  directions. In both examples $\lambda=1$, and  $\rho=0.1$ }
\end{figure}
(i) Let us first consider the simplest commensurate case $\theta=0$. The wave package spreads
 along the transverse direction, Fig.~\ref{figure3}a.
The mean square $\sigma_x^2$ presents small amplitude oscillations, while  $\sigma_y^2 $  shows a ballistic growing, Fig.~\ref{figure4}a.
(ii) Incommensurate orientation:  $\theta=\tan^{-1}\alpha_G$.   Fig.~\ref{figure3}b clearly shows that  the wave packet is dynamically  localized  in both directions, within a finite region of the lattice. In  Fig.~\ref{figure4}b  it is shown  that both  $\sigma_x^2$ and $\sigma_y^2$ oscillate. It is interestig to compare the  periods of these oscillations:
 the period of $\sigma^2_x$ is   $T_x \approx 6.9\tau$ and $\sigma^2_y$ has a period $T_y \approx11.2 \tau$.
The ratio of this two periods is similar to the golden ratio
$T_x/T_y\sim E_y/E_x=(\sqrt{5}-1)/2$,  which seems to be a consequence
of the relation between the two components of the drift velocity
$v_D=E\left(\sin\theta,-\cos\theta\right)/B$. Along the $x$ direction
the electron will take a time $T_x=qa/v_D\sin\theta$ to reach the
extreme of the unit cell, while on  the $y$ axis we have
$T_y=a/v_D\sin\theta$,  giving the correct relation between the
two periods. Each time the electron reaches the border of
a unit cell it reflects showing an oscillation in the mean square
dispersions. The oscillations are damped because of 
tunneling.

The comparison of the   commensurate (Fig.~\ref{figure3}a) with  incommensurate (Fig.~\ref{figure3}b) cases clearly  shows the very different transport properties of the electron depending on the relative  orientation of the electric field with respect to the lattice.  
The behavior of the   Shanon entropy is consistent with these results  (Fig.~\ref{figure4}c), we observe 
that for small times the evolution of $S$ is common for both cases, however for $ t \ge  3 \tau$ the growth of $S(t)$  corresponding to the irrational orientation is considerable  reduced as compared to the rational case.  
%

%FIGURA
\begin{figure}
\includegraphics[width=130mm]{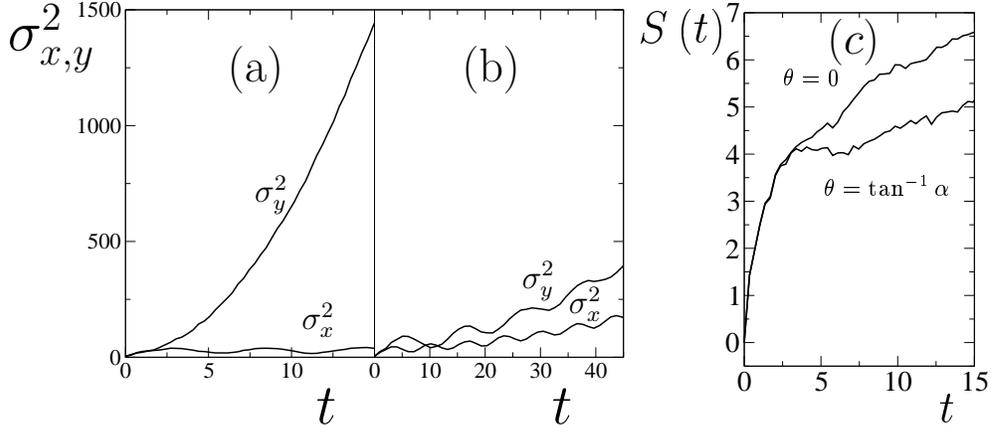}% Here is how to import EPS art
\caption{\label{figure4}
(a) Time evolution of the $\sigma_x^2$ and $\sigma_y^2$ variances when  the electric field is  aligned in
a commensurate direction $\theta=0$.   (b)  Time evolution of  $\sigma_x^2$ and $\sigma_y^2$ for an  incommensurate direction
$\theta=\tan^{-1}\alpha_G$ of $\mathbf{E}$.  In both examples  $\lambda = 1$, and $\rho = 0.1$. For the rational orientation of $\mathbf{E}$ the electron is only localized in the longitudinal direction, whereas  for $\theta=\tan^{-1}\alpha_G$ it is localized in both directions. (c) Time evolution for the  Shannon entropy for the two orientations of $\mathbf{E}$.}
\end{figure}

The localization for incommensurate directions of the electric field
can be understood with the following argument.  For the electric field orientation  $tan \, \theta = m_2/m_1$, the periodicity  is preserved  along the longitudinal and transverse direction but with an extended lattice of side dimensions 
$b = a \, \sqrt{m_1^2 + m_2^2} $. If additionally,   the number  of flux quanta in each of the new unit cells is given by a rational number, $i.e.$    $\sigma^{-1}= b^2 B/(h/e) = p/q$, then the  electric-magnetic  translation symmetries are preserved in the extended lattice of dimensions $q \, b \times b = q \, (m_1^2 + m_2^2) \, a^2$ \cite{Kunold1,Torres1}.  With   the new conditions,  the on-site energies of the original lattice are no longer the same,  in order to coherently propagate  the electron has to tunnel a distance proportional to the new lattice dimensions  $q \, (m_1^2 + m_2^2) \, a^2$. Clearly  as $m_1$ or $m_2$ increases the wave packet  diffusion will be inhibited. For irrational orientations of the electric field,  we can consider the rational approximant $tan \, \theta = m^{(n)}_2/m^{(n)}_1$,  then  the system becomes quasi-periodic   and the electrons are not  allowed to propagate coherently thus leading to localization.

We have analyzed  the time evolution of  electron  wave packets in  a  two-dimensional periodic  system in the presence of magnetic and electric fields. The dynamics  is governed by the effective Hamiltonian in (\ref{ele}), which consistently  includes  the coupling between Landau levels as well as the  periodic and external field contributions.
The inclusion  of an   electric  field (for commensurate orientations)  yields  localization in the 
longitudinal direction;  propagation   takes place along the transverse direction.  This effect   is a physical consequence  of the  quantization of the energy levels with respect to the longitudinal quasi-momentum. 
 Transition from extended to localized  states are  observed as   $\tan \, \theta$ changes from rational to irrational values.  It is expected that these phenomena should be experimentally accessible using lateral superlattices on semiconductor heterostructures, for example  by measuring assymetry effect on the magnetoresistance and Hall conductance as the electric field orientation with respect to the lattice axis is varied.


\begin{thebibliography}{00}


 \bibitem{Aubry1}
S. Aubry and C. Andre,
{\em  Ann. Israel Phys.  Soc. \/}
{\bf  3}  
(1980)
133. 

\bibitem{Harper1}
 P. G. Harper,
 {\em Proc. Phys. Soc. \/}
  {\bf A 68 }
  (1955)
874.

\bibitem{Rauh1}
A. Rau,
 {\em Phys. Status Solidi B \/}
  {\bf 69}
 (1975)
K9.

\bibitem{Peierls1}
R. Peierls,
{\em Z. Phys. \/}
 {\bf 80 } 
(1933)
  763.
 



\bibitem{Zak2}
  J. Zak,
 {\em Phys. Rev. Lett.}
 {\bf 79 }
(1997)
 533.

\bibitem{Hofta1}
D. R. Hofstadter,
{\em Phys. Rev. B \/}
{\bf 14 }
  (1976)
 2239.

\bibitem{Kunold1}
A. Kunold and M. Torres,
{ \em Phys. Rev. B \/}
 {\bf 61 }
(2000)
9879.

\bibitem{Torres1}
  A. Kunold and M. Torres,
{\em cond-mat/0311111 v1 \/}
 (2003).

\bibitem{Brito1}
 H. N. Nazareno and P. E. de Brito,
{\em Phys. Rev. B \/}
{\bf 64 }
 (2001)
045112.
 
\bibitem{Claro1}
  A. Barelli,  J. Bellissard and F. Claro,
 {\em Phys. Rev. Lett. \/}
{\bf 83 }
 (1999)
5082.
 
\bibitem{Esslin3}
T. Schl{\"o}sser, K. Ensslin,  J. P. Kotthaus and M. Holland,
{\em Europhys. Lett. \/}
  {\bf 33 }
 (1996)
683. 

\bibitem{Klit4}
C. Albrecht,  J. H. Smet,  D. Weiss,
          K. von Klitzing,  R. Hennig, M. Langenbuch,
          M. Suhrke, U. Rssler,  V. Umansky
          and H. Schweizer",
{\em Phys. Rev. Lett. \/}
{\bf 83 }
(1999)
2234. 

\bibitem{Klit2}
 C. Albrecht,  J. H. Smet, K. von Klitzing,  D. Weiss , V. Umansky and H. Schweizer,
{\em Phys. Rev. Lett. \/}
  {\bf 86 }
(2001)
 147.

\bibitem{Ashby1}
N. Ashby and S.C. Miller,
 {\em Phys. Rev. B \/}
  {\bf 139 }
 (1965)
A428. 

\bibitem{Shannon1}
  C. E. Shannon,
  {\em Bell Syst. \/}
  {\bf 109 }
 (195)
1492.
 
  \bibitem{Ketz5}
 R. Ketzmerick and K. Kruse and D. Springsguth  and T. Geisel,
 {\em Phys. Rev. Lett. \/}
 {\bf  84 }
(2000)
  2929.
  
 \end{thebibliography}
\end{document}